\documentstyle[pre,aps,psfig,epsfig,amssymb]{revtex}

\begin{document}
\title{\bf Discrete soliton ratchets driven by biharmonic fields}
\author{Yaroslav Zolotaryuk$^\dag$ and Mario Salerno$^\ddag$ }
\address{$^\dag$ Bogolyubov Institute for Theoretical Physics,
 National Academy of Sciences of Ukraine \\
Kyiv 03143, Ukraine \\
$^\ddag$ Dipartamento di Fisica ``E.~R.~Caianiello'' and
\\Consorzio Nazionale Interuniversitario per le Scienze Fisiche
della Materia (CNISM),\\ Universit\'{a} di Salerno, I-84081
Baronissi, Salerno, Italy}
\date{\today}
\wideabs{
\maketitle

\begin{abstract}
Directed motion of topological solitons (kinks or antikinks) in the
damped and AC-driven discrete
sine-Gordon system is investigated. We show that if the
driving field breaks certain time-space symmetries, the soliton can
perform unidirectional motion. The phenomenon resembles the  well
known effects of  {\it ratchet} transport and nonlinear harmonic mixing.
Direction of the motion and its velocity
depends on the shape of the AC drive. Necessary conditions for the
occurrence of the effect are formulated.
In comparison with the previously studied continuum case, the discrete
case shows a number of new features: non-zero depinning threshold for
the driving amplitude, locking to the rational fractions of the driving
frequency, and diffusive ratchet motion in the case of weak intersite
coupling.
\end{abstract}
}
 {Pacs: 05.45.-a, 07.90.+c, 89.20.-a}

\section{Introduction}
\label{intro}

Transport phenomena induced by the interplay between
non-equilibrium fluctuations, symmetry breaking and nonlinearity,
have recently attracted a great deal of interest. In particular,
point particle ratchets described by ordinary differential
equations have been largely investigated due to their relevance in
several fields, including  molecular motors and  Josephson
junctions (see reviews \cite{jap97rmp,r02pr}). In simple terms,
the point particle ratchets appear as the unidirectional motion of a
damped and driven particle, which is achieved under the influence
of only stochastic and deterministic forces of zero average,
independently on initial conditions. The phenomenon was ascribed
to the breaking of the symmetries connecting orbits with opposite
velocities in the phase space \cite{fyz00prl,yfzo01epl,dfoyz02pre}
and to the phase locking of the particle dynamics to the external
driver \cite{BS00,BS01}.

Ratchet phenomena in infinite-dimensional systems described by
nonlinear partial differential equations of soliton type have
also been investigated. This has been done for both spatially
asymmetric potentials with temporarily symmetric AC fields 
\cite{m96prl,SQ02} and
for symmetric potentials with temporarily asymmetric AC fields
\cite{sz02pre,fzmf02prl,zsc03ijmpb,goldobin01}. In both cases, the
ratchet effect appears as a unidirectional motion of the soliton,
which resembles the drift dynamics observed for point particle
ratchets; from here the name of {\it soliton ratchets} originates
\cite{SQ02}. The soliton ratchets  induced by asymmetrical external
fields have been implemented experimentally in long Josephson
junctions \cite{uckzs04prl} by means of {\it nonlinear harmonic
mixings}. This approach has been shown to be effective also in
other physical contexts \cite{mix} (for a detailed list of
references see the review paper \cite{r02pr}). The existence of
soliton ratchets in long Josephson junctions was also
experimentally demonstrated by using asymmetric magnetic fields
\cite{cc01prl} and spatially asymmetric currents
\cite{goldobin05}. From the theoretical point of view, symmetry
breaking conditions to generate soliton ratchets were discussed in
Ref. \cite{fzmf02prl}. The mechanism underlying soliton ratchets was
proposed in Ref. \cite{SQ02} for the case of a perturbed asymmetric
double sine-Gordon  equation driven by a symmetric AC driver
and extended in Ref. \cite{sz02pre} to the case of the damped
sine-Gordon (SG) equation with asymmetric AC fields. In both cases,
the phenomenon  was ascribed to the existence of an internal
oscillation on the kink profile which, in the presence of damping,
couples to the translational mode of the kink and produces
transport. The internal vibration was shown to be spatially
asymmetric, thus giving directionality to the motion, and phase
locked to the external force. This mechanism, also known as the
{\it internal mode mechanism}, has been confirmed for a number of
systems such as the asymmetric
double sine-Gordon equation with symmetric driver
\cite{SQ02,QSRS05} and the SG system with temporal asymmetric
forces \cite{sz02pre,molinaPRL03,willis04}.

In contrast to the continuous case, however, discrete soliton
ratchets have been scarcely investigated (some work on spatially
asymmetric discrete soliton ratchets has been done in
Refs. \cite{stz97pre,fmff04pd}). In this case, one can expect that the
presence of the Peierls-Nabarro barrier can strongly influence the
transport of discrete solitons. It is therefore of interest to
investigate the conditions under which the discrete soliton ratchets
can exist. The present paper is just devoted to this investigation. 
More precisely, we study the ratchet dynamics  induced by temporarily
asymmetric forces of zero mean on topological solitons (kinks and
antikinks) of the discrete sine-Gordon (DSG) system. This equation
models a number of physical systems such as arrays of Josephson
junctions, crystal dislocations or charge-density waves (see
\cite{fm96ap,bk98pr}). In particular, we investigate the
conditions for the occurrence of soliton ratchets and study the
dependence of the average soliton velocity on the system
parameters. A comparison with the results derived for continuous
soliton ratchets in Ref. \cite{sz02pre} is also provided.

From our study it emerges that discrete soliton ratchets are much
more complicated than their continuum counterpart. In particular,
the mean velocity of the kink in most cases appears to be a
piece-wise function of the parameters which resembles a devil's
staircase. We find that kink transport becomes very effective on
the corresponding orbits (limit cycles) which are phase locked to
the external driver. Transport is possible also in the presence of
more complicated dynamics such as chaotic and intermittency
orbits, especially when the system becomes very discrete (this is
achieved when the coupling constant is very small). In these
cases, however, the kink transport is not very efficient since the
drift velocity is rather small. Except for the very discrete case,
dominated by the pinning of the kink to the lattice, we find that the
internal mode mechanism remains valid also in the DSG ratchet in
all cases in which the transport is observed.

We remark that the soliton ratchets induced by temporarily asymmetric
fields may be an effective way to control the transport properties
of a large variety of continuous and discrete systems. From the 
experimental point of view, indeed, it is much easier to
produce ratchets by means of asymmetric fields than by breaking
the internal spatial symmetry of the system \cite{note1}, since in
the former  case no structural changes of the system are required.

The paper is organized as follows. In Section \ref{model}, we
present the model and derive the necessary condition for the
directed kink motion in terms of a simple symmetry analysis, which
is based on a point particle description of the kink dynamics. In
this section we
also describe the desymmetrization mechanism and confirm the
results by means of numerical simulations. In the next section, we
study the dependence of discrete kink ratchets on the system
parameters. In particular, we investigate the dependence of the
mean velocity of the kink on the amplitude, phase, and frequency
of the AC driver as well as on the damping and the coupling
constant. Moreover, the validity of the  internal mode
mechanism in the discrete case is discussed. In Section \ref{jja},
we consider the
soliton ratchet in a finite lattice and discuss possible
applications of the phenomenon to arrays of small Josephson
junctions. Finally, in Section \ref{conc}, we summarize the main
conclusions of the paper.

\section{The model}
\label{model}

The AC-driven and damped discrete sine-Gordon (DSG) equation is
introduced in a dimensionless form as follows
\begin{equation}
{\ddot u}_n-\;\kappa\; \Delta u_n\; +\; \sin u_n+\alpha {\dot
u}_n+E(t)=0, \;  n=1,2,\ldots N. \;
\label{1}
\end{equation}
Here $u_n$ is the displacement of the $n$th particle from its
equilibrium position, $\Delta u_n \equiv u_{n+1}-2u_n+u_{n-1}$ is
the discrete Laplacian,  $\kappa$ is the coupling constant
measuring the discreteness of the lattice, $\alpha$ is the damping
coefficient and  $E(t)$ is an external driving field. In the
following we assume $E(t)$ to be of the form
\begin{equation}
E(t)=E_1 \cos(\omega t)+E_2 \cos (m \omega t+ \theta)~, 
\label{2}
\end{equation}
where $m$ is an integer even number. Notice that the superposition
of two harmonics makes the periodic force to be asymmetric in
time for almost all values of $\theta$, a feature which can be
used to break the temporal symmetry of the system (see below). In
this context, it is of interest to investigate the condition under
which a driving force of zero mean of type (\ref{2}) can induce
kink's unidirectional motion similar to the one observed in the
continuum SG case \cite{sz02pre}. In this regard, we remark that
in the lower approximation, a discrete kink of the form
$u_n(t)=4\arctan {\left \{ \exp [n-X(t)]\right \}}$, can be viewed
as a single particle \cite{bk98pr} and its dynamics is described in
terms of collective coordinates: the center of mass $X(t)$  and
the kink velocity $\dot X(t)$. In this approach, the effective
point-particle equation of the motion becomes
\begin{equation}
{\ddot X}+\alpha {\dot X}+V'_{PN}(X)+{\tilde E}(t)=0 ~,
\label{PN}
\end{equation}
where $V_{PN}(X)=V_{PN}(X+1)$, $V'_{PN}(X) \sim \sin {2 \pi X}$ is
the Peierls-Nabarro (PN) potential accounting for the
discreteness of the lattice and ${\tilde E}(t) \sim E(t)$ is 
the effective driving field of the kink [we assume $\tilde E(t)$
to be proportional to $E(t)$]. An important parameter of the
problem is the frequency of kink oscillations in the bottom of
the PN potential (the PN frequency), which can be written as
\cite{im82jpsj}
\begin{equation}
\omega_{PN}=\sqrt{2\pi \alpha_0 \kappa^{3/2}}
\exp {(-\pi^2\sqrt{\kappa}/2)}
\;,\;\; \alpha_0 \simeq 30 \pi \;.
\label{PNfreq}
\end{equation}
Within this approximation, the unidirectional motion of the kink
corresponds  to a limit cycle of Eq.~(\ref{PN}), which is phase
locked to the frequency of the external driver. On this orbit, the
average kink velocity is expressed as
\begin{equation}
\langle v \rangle = \langle {\dot X}(t) \rangle= \frac{k}{l}\cdot
\frac{\omega}{2\pi} \;, \label{5a}
\end{equation}
with $k$ and $l$ being integer numbers. Notice that in this resonant
regime, the kink travels $k$ sites during the time $lT=2\pi
l/\omega$, so that, except for a shift in space, its profile is
completely reproduced after this time interval (in the pendulum
analogy, this orbit corresponds to $k$ full rotations of the 
pendulum during $l$ periods of the external drive). In the 
following, we
will refer to the phase locked dynamics also as to {\it resonances}.

\subsection{Symmetry properties and  conditions
for transport}

In analogy with the continuous SG case\cite{sz02pre,fzmf02prl},
one can expect that the directed kink motion arises when all the
symmetries of Eq.~(\ref{PN}), which relate kink solutions with
opposite velocities, are broken. Qualitative conditions for the
occurrence of this directed motion can be obtained from the 
analysis of the
symmetry properties of Eq.~(\ref{PN}). In this approach, the
many-particle problem is reduced to the one-particle ratchet
studied before \cite{fyz00prl,yfzo01epl} [we neglect oscillations
of the discrete kink profile (to be discussed later) which also
contribute to the phenomenon]. Our analysis is based on the simple
observation that the sign of the soliton velocity ${\dot X}(t)$
can be changed by means of the following symmetry operations:
\begin{eqnarray}
&& \label{5b} {\bf \hat D_{\cal X}}: t \rightarrow t+T/2, \; X
\rightarrow - X ,\\
&&
 {\bf \hat D_{\cal T}}: X
\rightarrow X + X_0, \;   t \rightarrow -t + 2t_0~, \label{5bb}
\end{eqnarray}
where  ${\bf \hat D_{\cal X}}$ denotes a shift in time followed by
a reflection is space and ${\bf \hat D_{\cal T}}$ is a shift in
space followed by a reflection in time (here $t_0$ is a constant
and $X_0$ is either an integer or a half integer). Notice that
Eq. (\ref{PN}) is always invariant under the symmetry ${\bf \hat
D_{\cal X}}$ provided the external driver satisfies
\begin{equation}
E(t)=-E(t+T/2)\;. \label{5}
\end{equation}
Notice that $V_{PN}(-X)=V_{PN}(X)$ is always satisfied because
the sine-Gordon potential is symmetric. In the zero damping limit
($\alpha\rightarrow 0$), Eq.~(\ref{PN}) becomes invariant also
under the symmetry ${\bf \hat D_{\cal T}}$ with $X_0=1$, provided
the external driver satisfies the condition
\begin{equation}
E(t+t_0)=E(-t+t_0) \;, \label{5c}
\end{equation}
with $t_0$ being a constant, which depends on the shape of $E(t)$. In 
the
overdamped limit ($\alpha \rightarrow \infty$), Eq.~(\ref{PN}) can
be rewritten as ${\dot X}+V'_{PN}(X)+{\tilde E}(t)=0$ from which
one can see that it becomes invariant under the symmetry ${\bf \hat
D_{\cal T}}$ with $X_0=1/2$, provided the external driver
satisfies the condition
\begin{equation}
E(t+t_0)=-E(-t+t_0)\;. \label{9}
\end{equation}
It should be remarked here that $V'_{PN}(X)=-V'_{PN}(X+1/2)$ is always 
satisfied since for the DSG equation we have that 
$V'_{PN}(X) \sim \sin 2\pi X$). From the above symmetry properties 
it follows that one can
break all symmetries relating orbits with opposite velocities by
properly choosing the driving force $E(t)$. Thus, for the general
case $\alpha \neq 0$, we have only the symmetry ${\bf \hat
D_{\cal X}}$ for  Eq. (\ref{PN}). This symmetry  can be broken by
choosing any function $E(t)$ which violates Eq.~(\ref{5}). In the
zero damping limit ($\alpha\rightarrow 0$), we have, besides ${\bf
\hat D_{\cal X}}$, also the symmetry ${\bf \hat D_{\cal T}}$ with
$X_0=1$. In this case, one must choose a driving field $E(t)$ which
violates both Eq.~(\ref{5}) and Eq. (\ref{5c}). Similarly, in the
overdamped limit ($\alpha \rightarrow \infty$), a function $E(t)$
which violates both Eq. (\ref{5}) and Eq. (\ref{9}) should be
chosen.

From these considerations it follows that a simple sinusoidal
driver cannot support the kink transport in the lattice, since Eqs.
(\ref{5b})-(\ref{9}) in this case are always satisfied. For a
biharmonic driver of the type (\ref{2}), however, we have that 
Eq.~(\ref{5}) is violated for any $\theta$ (if $m$ is even), so that
the kink transport should become possible.

%
%
\begin{figure}[htb]
\vspace{2pt}
\centerline{\psfig{file=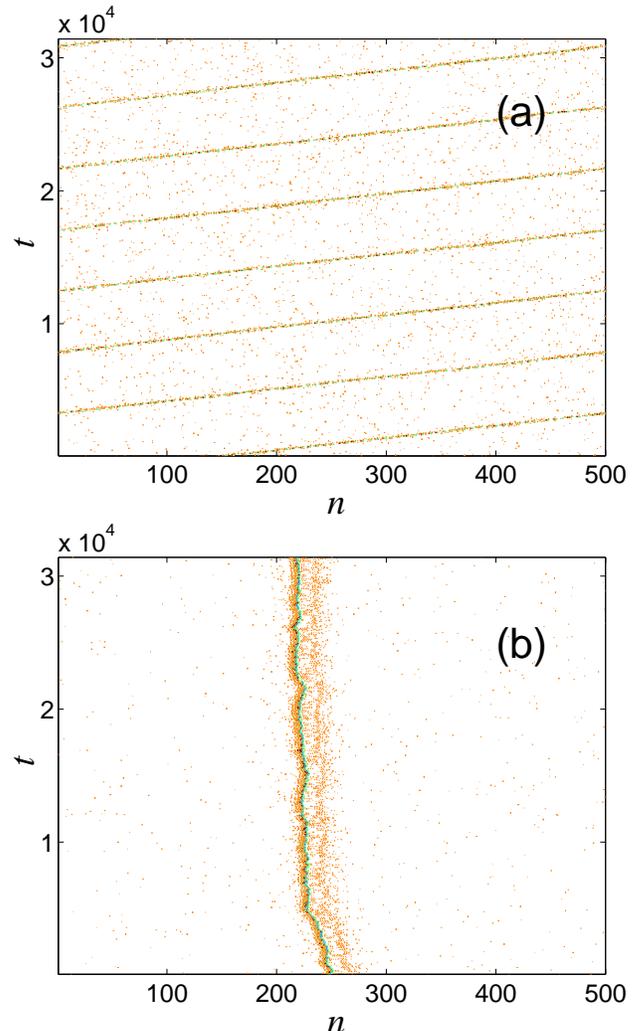,width=3.3in,angle=0}}
\caption{Contour plot of the temporal evolution of the particle
velocities ${\dot u}_n(t)$ for (a) $E_1=0.3$, $E_2=0.15$ and (b)
$E_1=0.45$, $E_2=0$. Other parameters are:
 $\kappa=1$, $\alpha=0.05$, $\omega=0.1$, $\theta=2$, $D=0.002$,
$N=500$. Periodic boundary conditions have been applied.}
\label{fig1}
\end{figure}

From these arguments it is also clear that an external periodic 
driver of
zero mean, which consists of the superposition of only the first
two harmonics, is the simplest driver that can be used to induce 
the soliton ratchets in the DSG system.

\subsection{Numerical study of transport vs symmetries}
%

To verify  the validity of the previous analysis and to check the
desymmetrization of the orbits as a function of the driver
parameters $E_1, E_2$, and $\theta$, we recourse to direct numerical
integration of Eqs.~(\ref{1}). In order to be sure that the system
explores the whole phase space and that the phenomenon does not
depend on initial conditions, we perform the first step
simulations in the presence of white noise.
%
\begin{figure}[htb]
\vspace{2pt}
\centerline{\psfig{file=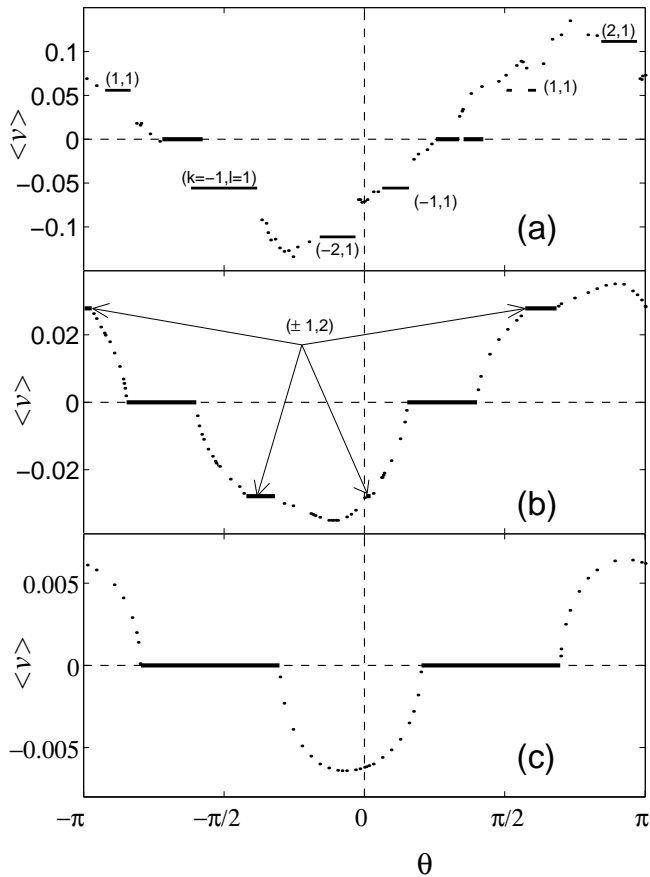,width=3.5in,angle=0}}
\caption{Dependence of the average kink velocity on the phase
difference $\theta$ for $E_1=E_2=0.2$, $\alpha=0.05$ (a),
$\alpha=0.2$ (b) and $\alpha=0.5$ (c). Other parameters are:
$\kappa=1$, $\omega=0.35$. Corresponding pairs of rotation numbers
$(k,l)$ are given nearby the most pronounced resonances. Dashed lines
mark the coordinate axes.} 
\label{fig2}
\end{figure}
White noise has been included by adding in the r.h.s. of Eqs.~(\ref{1}) 
a stochastic term $\xi_n(t)$ of zero mean, $\langle
\xi_n(t) \rangle =0$, and with the autocorrelation function $\langle
\xi_n(t)\xi_m(t')\rangle= 2 \alpha D \delta_{mn}\delta (t-t')$.
The resulting Langevin equations have been integrated numerically 
by using the fourth-order Runge-Kutta method, adopting either 
free ends
or periodic boundary conditions: $u_{n+N}(t)=u_n(t) \pm 2\pi$
(positive and negative signs refer to kinks and antikinks,
respectively).

 Figure \ref{fig1} illustrates the dynamics of a kink of the damped
DSG equation under the influence of noise, driven by a biharmonic driver
(\ref{2}) with $m=2$ [panel (a)], and by a single harmonic driver
[panel (b)]. We see that while the single harmonic driver is
unable to produce directed motion, the biharmonic driver is quite
effective to produce the kink transport. From this figure, one can
also
observe that there is only one attractor corresponding to the
unidirectional motion of the kink. Further investigations (see
below and Sec. IV) show that this is true in almost all cases,
except for very narrow intervals in the parameter space where two
attractors can coexist.
%
%
\begin{figure}[htb]
\centerline{\psfig{file=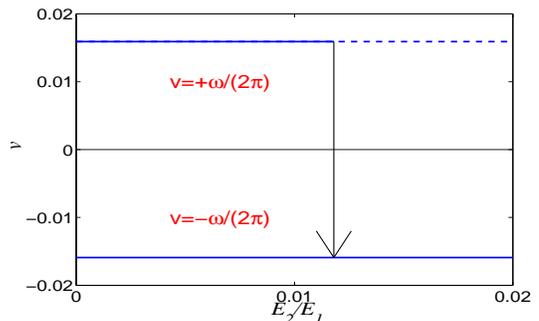,width=2.8in,height=1.7in,
angle=0}} \caption{Velocities of two kinks, running into the
opposite directions at $\alpha=0.05$, $\omega=0.1$, $E_1=0.21$,
$\theta=0$ as a function of $E_2/E_1$. Dashed line denotes that
the given solution is no longer stable.} 
\label{fig3}
\end{figure}
This fact makes possible to investigate Eqs.~(\ref{1}) in the absence
of noise  without averaging over the initial conditions. 
In Fig.~\ref{fig2} we plot the time average kink velocity $\langle v
\rangle$ as a function of the phase difference $\theta$ in the
case of a biharmonic driver with $m=2$. In contrast to the
continuous case  for which $\langle v \rangle$ was shown to have a
sinusoidal dependence on $\theta$ \cite{sz02pre,zsc03ijmpb}, in
the discrete case, we find a complicated piece-wise dependence
$\langle v \rangle (\theta)$, which resembles a sinusoidal function
only slightly. This is due to the fact that in the most of cases the
dynamics is phase locked to the external driver and the kink
velocity is given by Eq.~(\ref{5a}).

Notice that for a weak damping [see panel (a) of Fig.~\ref{fig2}]
several resonances ($k=l=\pm 1$, $k=\pm 2,l=1$) are clearly
visible, while for a strong damping [panel (c)], the kink becomes
pinned to the lattice for large intervals of $\theta$ (depinning
of the kink would require stronger fields). From 
Figs.~\ref{fig2}(a-c) a qualitative understanding of the symmetries 
to break, in order to achieve unidirectional motion, can be obtained.
In particular, one can see that close to the underdamped limit
($\alpha=0.05$), the mean velocity becomes zero  in the intervals
$\theta \in [-2.26,-1.81]$, $\theta \in [0.80,1.06]$, and  $\theta
\in [1.11,1.33]$ [see Fig.~\ref{fig2}(a)]. For larger values of the
damping constant [see Fig.\ref{fig2}(b)], we have  
$\langle v\rangle=0$ in the intervals $\theta \in [-2.66,-1.88]$ and
$\theta \in [0.48,1.26]$, while for $\alpha=0.5$, the kink stays
pinned for $\theta \in [-2.5,-0.95]$ and $\theta \in [0.64,2.19]$.
By increasing $\alpha$, the extremal values of $\langle v \rangle$
shift to $0$ or $\pm \pi$ with the growth of $\alpha$. Moreover,
the length of the intervals in $\theta$ where $\langle v \rangle=0$ 
increases, while the average kink velocity decreases as
$\alpha$ increases. This behavior is an obvious consequence of the
slowing down effect of the damping on the kink motion. In the
overdamped limit $\alpha \rightarrow \infty$, the intervals become
centered around the points $\theta=\pm \pi/2$ for which $E(t)=E_1
\cos \omega t +E_2\cos (2\omega t+{\pi}/{2} )=-E(\pi/\omega-t)$
and thus the symmetry (\ref{5b}) is satisfied. Instead, in the
underdamped case ($\alpha \rightarrow 0$), the centers of
these intervals gradually shift to the positions $\theta= 0,\pm
\pi$ for which $E(t)=E(-t)$ and the symmetry (\ref{5bb}) is
satisfied.

We recall that in the continuum case \cite{sz02pre}, the functional
dependencies obtained in the underdamped and overdamped limits were
found to be $\langle v \rangle \sim \sin \theta$ 
and $\langle v \rangle \sim
\cos \theta$, respectively. Thus, the transition from one limit to
another in the discrete case is similar to the continuous case in
the sense that the directed soliton motion disappears in
correspondence with values of $\theta$ for which the respective 
symmetry is restored.
%
\begin{figure}[htb]
\vspace{2pt}
\centerline{\epsfig{file=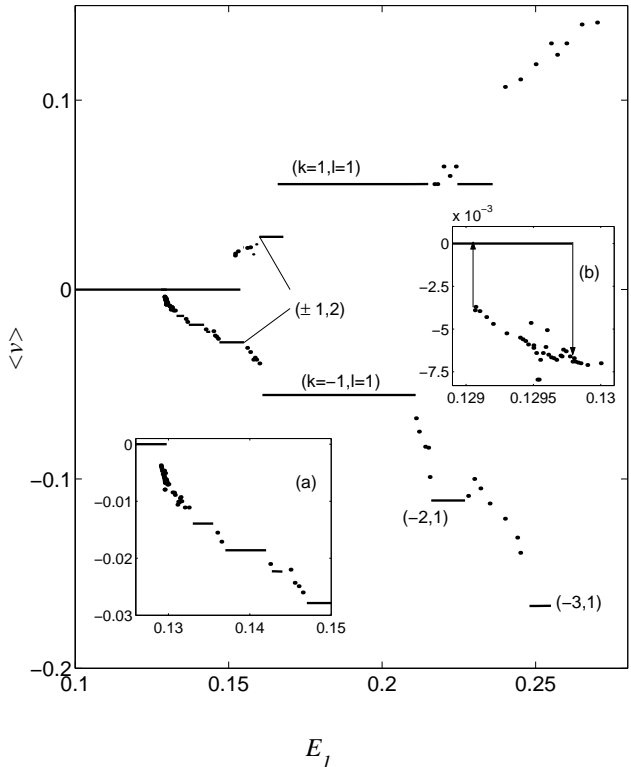,width=3.3in,angle=0}}
\caption{Dependence of the average kink velocity on the amplitude
of the driver $E_1=E_2$ for $\theta=-0.5$ (decreasing dependence) and
$\theta=2$ (increasing dependence). Other parameters are: $\alpha=0.1$,
$\kappa=1$, $\omega=0.35$. Numbers in brackets show the
respective pair of rotation numbers $(k,l)$. Insets show more
details for case $\theta=-0.5$.} 
\label{fig4}
\end{figure}
In the discrete case, however, the soliton velocity is zero not
only for those values of $\theta$ which restore the symmetries 
(\ref{5b}) and (\ref{5bb}), but also
for some finite interval around these values.

In the case of odd values of $m$, the condition (\ref{5}) is
satisfied for any value of $\theta$ if $E_2\neq 0$. We have
indeed that $E(t+T/2)= E_1\cos (\omega t+\pi)+E_2\cos (m\omega
t+m\pi+\theta)=-E(t)$ and therefore the symmetry ${\bf \hat
D_{\cal X}}$ is always valid. This observation implies that there 
should be no kink transport in the system, a result which is 
indeed confirmed
by direct numerical simulations of the full system with $m=3$,
both in the presence and in the absence of noise. Also, in 
analogy with
the continuous case, we find that for the same parameter values
the discrete antikink ratchets always move in the direction 
opposite to the kink motion.

The emergence of the directed kink motion can also be seen as a
consequence of the desymmetrization of the basins of attraction of
the two limit cycles corresponding to kinks moving with opposite
velocities (for single particle ratchets, the desymmetrization of
the orbits has been shown in Refs. \cite{fyz00prl,bs00pre}).
%
%
\begin{figure}[htb]
\vspace{2pt} \centerline{\epsfig{file=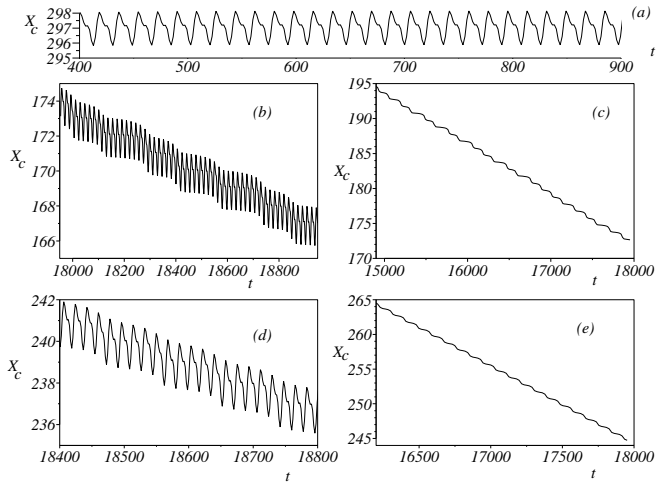,
height=3.6in,angle=-90}} \caption{Coordinate of the kink center
$X_c$ as a function of time for different values of the driving
amplitude $E_1=E_2=0.129$ (a); $0.12981$ (b,c)  and $0.1309$(d,e).
In panels (c) and (e), $X_c(t)$ has been plotted after each
oscillation period $T$ (see text for details). Other parameters
are: $\alpha=0.1$, $\omega=0.35$, $\kappa=1$, $\theta=-0.5$.}
\label{fig5}
\end{figure}
As soon as the symmetry is broken (by switching on the field
$E_2$), one of the basins of attraction begins to shrink and
eventually disappears as $E_2$ increases. As a result, only the
limit cycles which correspond to motion in one direction survive,
as one can see from Fig. \ref{fig3}. In particular, for the
parameters given in this figure, we find that the attractor
corresponding to the limit cycle with rotation numbers $k=1,l=1$
disappears at $E_2/E_1=0.0118$, i.e., already for a rather weak
asymmetry of the field ($E_2\ll E_1$).

\section{Dependence of the phenomenon on system parameters}
\label{param}
In this section, we investigate the dependence of kink transport on
the system parameters and the internal mode mechanism.

\subsection{Dependence on the driving amplitude}

For continuous ratchets it was shown previously \cite{sz02pre} that the
average kink velocity is proportional to $E_1^2E_2$, so that,
provided the respective symmetries are broken,  the directed
motion can occur for arbitrary small values of the driving
amplitudes. For the DSG equation, the dependence of the average
kink velocity on the driver amplitudes is depicted in 
Fig.~\ref{fig4} from which one can see that the kink velocity is not a
smooth monotonic function of $E_{1,2}$, but a piecewise function
with plateaux of different lengths, resembling a ``devil's
staircase'' [the same as in Fig. \ref{fig2}(a) but monotonic]. 
The plateau values of the kink
velocities are given by Eq.~(\ref{5a}) and correspond to dynamical 
regimes,
which are limit cycles with rotation numbers $(k,l)$ phase-locked
to the driver. Notice that the largest resonant step ($\langle v
\rangle=0.0557$) is achieved at the rotation numbers corresponding 
to the main resonance ($k=\pm 1,l=1$). One can easily observe the 
smaller resonant steps with $k=\pm 2,l=1$ 
and  $k=\pm 1,l=2$, for which the kink velocities equal
$\langle v \rangle=0.1114$ and $\langle v \rangle=0.02785$,
respectively. Higher order resonances can also be identified, but
they are not well visible in the figure.

The main feature of the discrete case is the existence of a
threshold in the driver amplitude (depinning threshold), above
which the transport can occur and below which the kink is pinned to
a lattice site, oscillating around its center of mass:
\begin{equation}
X_c= \frac{\sum_{n=1}^N n(u_{n+1}-u_{n-1})}{2(u_N-u_1)}.
\label{10}
\end{equation}
In Fig.~\ref{fig5}, the kink position $X_c$ is depicted as a function 
of time for
different values of the driving amplitude. We see that when the
driving amplitude is below threshold [panel (a)], the kink
performs periodic oscillations around its center of mass, while
above the threshold [panels (b)-(e)], the standing kink becomes
unstable and the directed motion starts. By further increasing 
the
driving amplitude above the threshold, the kink motion becomes
either phase locked to the external driver or chaotic with an
intermittent behavior [see panels (b) and (c)]. Panels (d) and
(e) of Fig.~\ref{fig5} show the phase-locked dynamics with rotation
numbers $k=1,l=5$. Notice that in this case, the kink travels over
a fixed number of sites during each period $T$ in agreement with 
Eq.~\ref{5a}. By further increasing $E_1$, the  the kink dynamics
starts to switch between periodic (or quasiperiodic) regime and
intermittency. Similar transitions in the case of a
single harmonic driver have also been reported in
Refs.~\cite{fm96ap,mfmfs97prb}. In the intermittency regime, the 
kink dynamics switches in an unpredictable manner between two
attracting limit cycles [see panels (b) and (c) of
Fig.~\ref{fig5}]. In this case, the kink dwells some time interval
on a certain site, pinned by the corresponding minimum of the PN
barrier, before jumping to the next site. From panel (b) one can 
also see that the dwelling time is not rationally related to the
driving period $T$ and changes randomly from jump to jump. Thus
in the intermittent regime the kink does
not travel the same number of sites during an integer number of
periods of $E(t)$ and the global dynamics is chaotic.

The same behavior is seen in panel (c) of Fig.~\ref{fig5} for a
larger time scale. Note that the  intermittency behavior occurs
not only around the depinning threshold, but also for larger values
of the driving amplitude. Also notice, from the inset (b) of 
Fig.~\ref{fig4}, that the depinning threshold has an hysteretic
behavior. In this regard, we remark that the numerical
investigation has been performed by increasing the driving
amplitude in small steps, taking the final state for a given step
as an initial condition of the next one. When the pinned state loses
stability, the system finds itself on a chaotic attractor, which
corresponds to the directed motion. By increasing further $E_1$ we
obtain larger values of the kink velocity, while if we move
backwards, we observe that two attractors can coexist - one of them
corresponding to a pinned kink oscillating around its center of
mass, and the other to a kink performing ratchet dynamics. If
$E_1$ is further decreasing, the moving state loses stability and
the system jumps back to the pinned state. The width of the
hysteresis (interval between two bifurcations) appears to be
rather small, i.e., $0.129065<E_1<0.12979$. Similar hysteretic
phenomena also appear at larger values of the driver amplitudes.
For example, one can find that in the interval $0.1660<E_1<0.1678$, two
limit cycles with $k=1,l=2$ and $k=1,l=1$ coexist. A more
detailed numerical investigation of the $\langle v \rangle(E_1)$
dependence shows that the observed devil's staircase is incomplete
and there are gaps inside it, which depend on the value of
$\theta$. As the $\langle v \rangle(E_1)$ dependence becomes 
steeper, less phase locked states are found. In particular for the
case $\theta=2$, we see that in the transition from the $k=1,l=2$
to the $k=1,l=1$ state, all intermediate rational values of $k$
and $l$ are missing.

\subsection{Dependence on the driving frequency}

The dependence of the mean kink velocity on the driver frequency
was investigated in Ref.~\cite{sz02pre} for the continuous SG model. 
In this case, it was shown that the average velocity depends on the
driving frequency as $\langle v \rangle \sim \sin \left
[\theta-\theta_0(\omega;\alpha)\right] /\omega^3$, where $\tan
\theta_0(\omega;\alpha)=[\alpha/(2\omega)]
\cdot[3+(\alpha/\omega)^2]$. This fact implies that $\max_{\theta \in
[0,2\pi[} \langle v \rangle $ decays with the growth of $\omega$.
For a fixed value of $\theta$, however, this dependence is defined
by the mutual relations between $\omega$ and $\alpha$, so that
$\langle v \rangle$ can experience oscillations and sign reversals
before tending to zero either in a decreasing or in increasing
way. Similar behavior is expected also for the  DSG equation,
although in this case, the problem is complicated by the presence
of the Peierls-Nabarro (PN) frequency $\omega_{PN}$ in Eq.
(\ref{PN}). In analogy with Fig.~\ref{fig2}, one can expect the
dependence $\langle v \rangle(\omega)$ to be also a piecewise
function with a ``devil's staircase'' character.

\begin{figure}[htb]
\vspace{2pt} 
\centerline{ \epsfig{file=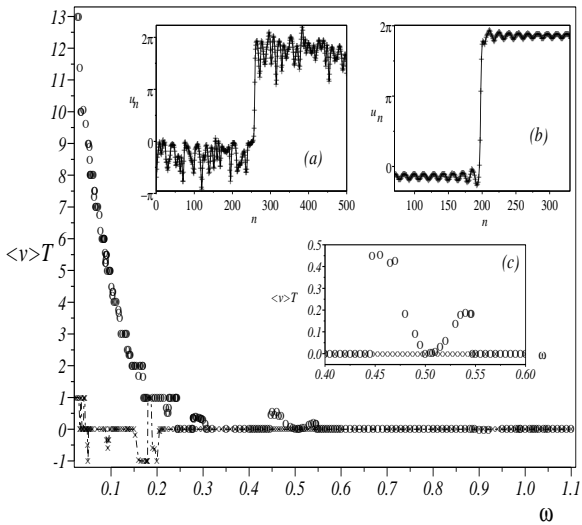,width=3.5in,
height=3.5in,angle=-90}} \caption{Dependence of the normalized
kink velocity $\left <v \right>T$ on the driving frequency for
$E_1=E_2=0.1$, $\theta=2$, $\alpha=0.05$ ($\circ$) and
$E_1=E_2=0.02$, $\theta=1.5$, $\alpha=0.15$ ($\times$). In both
cases $\kappa=1$.  The insets (a) and (b) show the kink profile
for the first set of parameters when $\omega=0.45$ and $\omega= 0.507$,
respectively. Inset (c) shows more detailed behavior around
$\omega=0.5$. } \label{fig6}
\end{figure}

In Fig.~\ref{fig6}, we depict the kink velocity, normalized to the
driving frequency ($\langle v \rangle T=2\pi v/\omega$) as a
function of $\omega$ for two values of the damping constant
(notice that for phase locked dynamics, $\langle v \rangle T$
coincides with the ratio of rotation numbers $k/l$). One can see 
that
the dynamics is characterized by a series of resonances, the most
pronounced one being at the main frequency ($k=1,l=1$). We also
observe that the resonances with $k>l$ are less pronounced and 
the subharmonic resonances with $k<l$ are practically not visible 
in the figure. The case of
small damping and larger driver amplitude (depicted in the figure
by circles) is characterized by an almost monotonic decay, while
in the opposite case (larger damping and smaller driving
amplitude), there are wide pinned regions and some peak around the
value of $\omega_{PN}$ [notice from Eq.~(\ref{PNfreq}) that
$\omega_{PN}\simeq 0.17$ for $\kappa=1$].
In both the cases we find that, except for the resonances  $\omega
\approx \omega_L$ and $2 \omega \approx \omega_L$,  where
$\omega_L$ is the frequency of linear waves
\begin{equation}
\omega_L^2(q)=2\kappa (1-\cos q)+1, \;
\end{equation}
the kink transport becomes effective mainly at low frequencies
$\omega \ll \omega_L$ and disappears for driving frequencies
$\omega > \omega_L$. Also, from Fig. \ref{fig6} one can see that the
kinks become pinned to the lattice for values of $\omega$ 
significantly smaller than $\omega_L$. At the
resonances $\omega \approx \omega_L$ and $2 \omega \approx
\omega_L$, we observe that the kink dynamics become coupled to 
linear waves in the system (plasmons in the case of array of
Josephson junctions). Insets (a) and (b) of Fig.~\ref{fig6} show
the profiles of the kink solution in these resonant cases, while
inset (c) shows details of the $\langle v \rangle$ dependence in
the neighborhood of the second resonance. The coupling of the kink
motion with linear waves occurs in the interval
$0.447<\omega<0.517$.  At the beginning of this interval, the kink
displays a chaotic tail as shown in Fig.~\ref{fig6}(a). As 
frequency increases, the chaotic tail becomes more and more
regular. This is shown in inset (b) of the figure, from which
one can see that at $\omega=0.507$ the oscillating tail is almost
monochromatic. At $\omega=0.517$ the oscillating tail turns into a
localized oscillating mode which decays at infinity. A further
increase of the driving frequency causes the decay of the width of
the mode.
%
%
\begin{figure}[htb]
\vspace{2pt} 
\centerline{\psfig{file=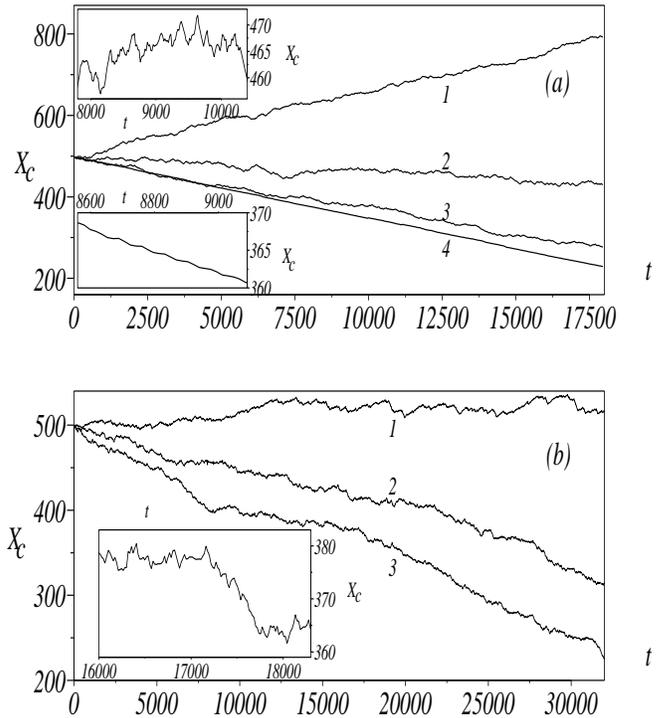,
width=4.4in,height=3.7in,angle=-90}} 
\caption{Time evolution of
the position of the kink center for: (a) $\kappa=0.5$;
$E_1=E_2=0.21$ (curve 1), $0.19$ (curve 2), $0.17$ (curve 3),
$0.11$ (curve 4) and (b) $\kappa=0.25$; $E_1=E_2=0.14$ (curve 1),
$0.24$ (curve 2), $0.22$ (curve 3). Other parameters are:
 $\theta=1.5$, $\alpha=0.1$, $\omega=0.35$.
The upper inset of panel (a) corresponds to details of the case
$E_1=E_2=0.19$ and the lower inset corresponds to the case
$E_1=E_2=0.11$. The inset of panel (b) corresponds to the case
$E_1=E_2=0.22$. In all figures, the data have been plotted with the
interval $T=2\pi/\omega$.} 
\label{fig7}
\end{figure}

A similar scenario occurs also at $\omega \approx \omega_L$, but in
this case the kink velocity is much smaller and practically not
visible in Fig.~\ref{fig6}. Notice that in the frequency interval
$0.895<\omega <1.05$, the kink becomes again coupled to linear modes and
displays a chaotic tail, which becomes more regular as $\omega$ 
further increases. At the beginning of this window, the dynamics
in the tails are strongly chaotic and they are accompanied by the
formation of large-amplitude localized excitations (breathers).

By reducing the driving amplitude, the frequency windows, in which
the coupling with linear waves occurs, decrease. We find that for
$E_1=E_2=0.04$, the first coupling window occurs at
$0.480<\omega<0.495$ and the second one at $0.955<\omega<0.994$.
In this case, no chaotic dressing of the kink is observed [the
coupling occurs with very few (or single) linear modes and the
kink looks very similar to that in Fig.~\ref{fig6}(b)]. For the
driving amplitude $E_1=E_2=0.01$, the coupling with linear waves
does not take place at all, and around the resonant frequencies
$\omega_L$ and $\omega_L/2$, the kink remains pinned to the
lattice. As the driving amplitude increases, beyond a certain
threshold chaotic oscillations can completely destroy the kink.
From these results, we conclude that the coupling of the kink dynamics
with linear waves depends very much on the amplitude of the
external field and this effect is nonlinear in $E_1$. Near the 
resonances with
linear modes, the kink dynamics become more chaotic (diffusive),
resembling the one of a Brownian particle (the kink makes many
random jumps backward and forward and the unidirectional  motion
can be seen only at a large time scale).

\subsection{Dependence on the coupling constant}
\label{kappa}

In this section, we investigate the dependence of the ratchet
phenomenon on the interaction constant $\kappa$ and the conditions to
sustain mobile kinks in the  system. As well known 
\cite{pk84pd}, the discreteness usually prevents free topological
soliton propagation. In our
model, the discreteness of the system is characterized by the
interaction constant $\kappa$, with $\kappa \rightarrow \infty$
corresponding to the continuum limit. In Fig.~\ref{fig7}, we depict
the temporal evolution of the kink center of mass $X_c$ for two
values of the coupling constant $\kappa=0.5$ [panel (a)] and
$\kappa=0.25$ [panel (b)]. From this figure, we conclude that the
unidirectional motion exists also for small values of $\kappa$
with dynamics which is chaotic rather than phase locked (see
insets of the figure).

Chaotic motion, leading to transport is found to be either of the
intermittent type [curve 4 of panel (a)] or diffusive type
[curves 1-3 of panel (a)]. The intermittent regime is more
often observed for $\kappa \gtrsim 1$ and the behavior is the
same as described in the previous subsection, i.e., the kink
oscillates in the minima of the PN potential until it jumps to the
adjacent site [see lower inset of panel (a)]. The motion is 
characterized by the fact
that the jumps of the kink occur randomly in time,
but always in the same direction. For $\kappa \lesssim 1$ the
diffusive motion is the most typical scenario. In this case, the
kink jumps randomly forward and backward [see upper inset of panel
(a) and inset of panel (b)], but on average the motion remains
unidirectional. A decrease of the coupling constant makes the
dynamics even more diffusive in the sense that the forward and
backward jumps become of larger amplitude with a consequent
decrease of the average velocity.

\begin{figure}[htb]
\centerline{\psfig{file=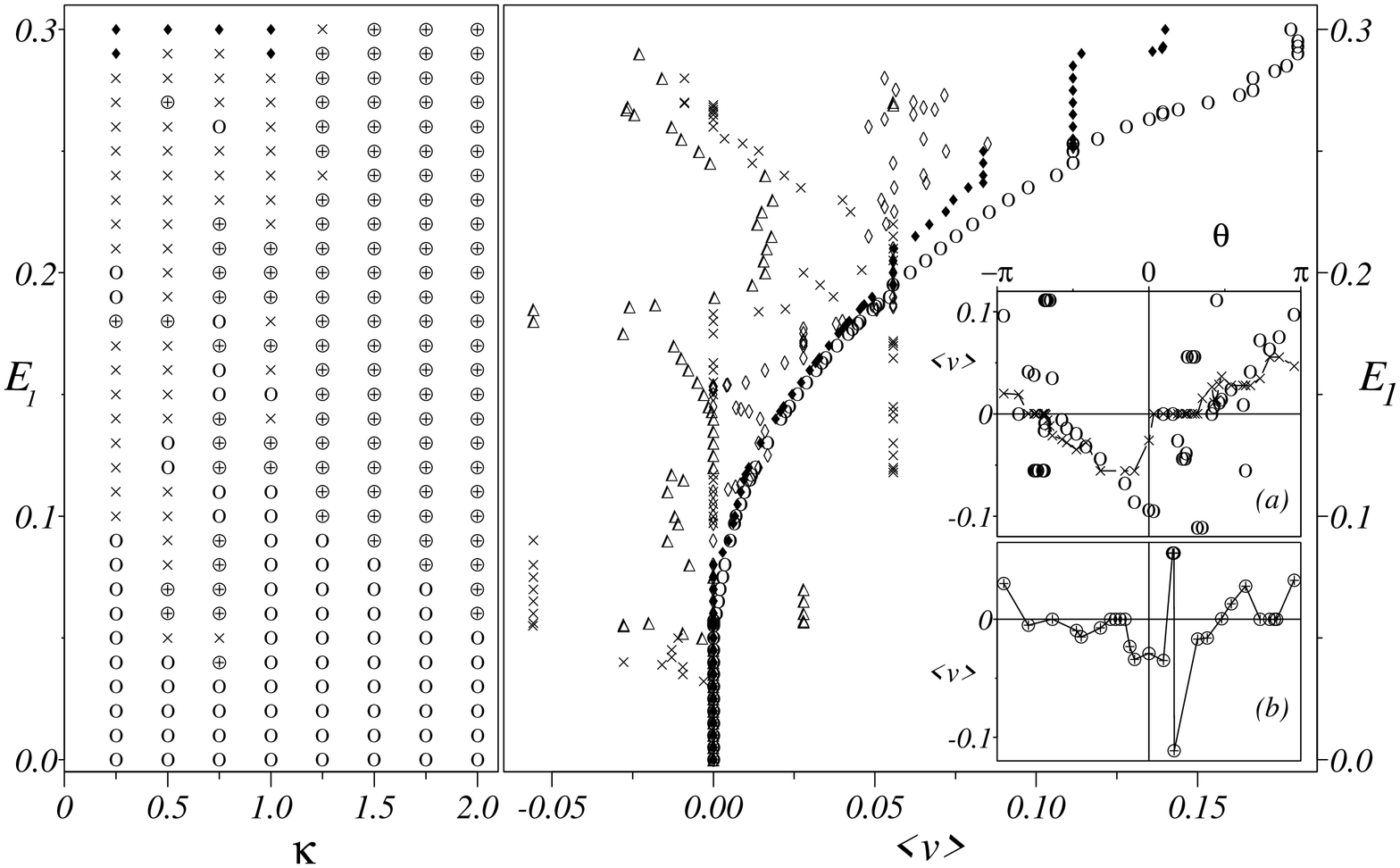,
width=3.3in,height=3.3in,angle=-90}} 
\caption{{\it Left panel.}
Diagram that demonstrates main regimes of the kink ratchet motion
for different values of the coupling constant $\kappa$ and the
driving amplitude  $E_1=E_2$ (other parameters are: $\theta=1.5$,
$\alpha=0.1$, and $\omega=0.35$). In the diagram $\circ$ stands for
the regime when kink is pinned ($\langle v \rangle =0$), $\oplus$
stands for the periodic or quasiperiodic ratchet motion, $\times$
stands for the chaotic regime, and $\blacklozenge$ corresponds to
the case when kink solutions do
not exist.\\
{\it Right panel.} Dependence of the average kink velocity on the
driving amplitude $E_1=E_2$ for different values of the coupling
constant
 $\kappa=0.5$($\Delta$), $\kappa=0.75$ ($\times$), $\kappa=1$ ($\diamond$),
$\kappa=1.5$ ($\blacklozenge$), and $\kappa=2$ ($\circ$). Other
parameters are as in the left panel. The insets show dependence of
the average velocity on the phase difference $\theta$. Inset (a)
corresponds to $\kappa=0.75$, $E_1=E_2=0.19$ ($\circ$) and
$E_1=E_2=0.25$ ($\times$), inset (b) corresponds to  $\kappa=0.5$,
$E_1=E_2=0.22$ ($\oplus$). Solid lines are drawn  as a guide to an
eye.} 
\label{fig8}
\end{figure}

In the left panel of Fig.~\ref{fig8} the diagram of possible
dynamical regimes in the plane $(\kappa, E_1)$ is shown. We
observe
that by changing the coupling constant $\kappa$, one can pass from
regular (phase locked) dynamics to chaos. The regular dynamics are
dominant for $\kappa > 1$ and when $\kappa$ is decreases, the windows
of chaotic motion appear. By further decreasing $\kappa$, the
number of chaotic windows increases and the chaotic motion becomes
dominant. At $\kappa=0.25$, almost no regular dynamics exist.
Notice that the diagram in Fig.~\ref{fig8} does not show the
complete picture and the details of transitions to different
dynamical regimes because in order to limit the computation time, we have
used in the numerical calculations the step of $0.01$ in the
amplitude of the driving field $E_1=E_2$ and a coarse step of
$0.25$ in $\kappa$. We find that for driving amplitudes $E_1=E_2
\gtrsim 0.3$, the chaotic dynamics of the {\it whole lattice}
destroy kink solutions.
%

In the right panel of Fig.~\ref{fig8}, we show the dependence of
the average kink velocity on the driving amplitude $E_1=E_2$ for
different values of the coupling constant $\kappa$. 
We observe that already for
$\kappa=2$, the curve is very smooth (except for the small
resonant steps $k=3,l=1$ and $k=3,l=2$) and the behavior becomes 
very similar
to that reported for the continuous limit ($\langle v \rangle
\sim E_1^2 E_2$) \cite{sz02pre}. By increasing the driving
amplitudes $E_{1,2}$, the dependence becomes non-monotonic, a fact
which was also observed in the continuous SG case and ascribed to
the interaction of the kink with internal oscillation modes
\cite{sz02pre}. On the other hand, decrease of the coupling
constant makes the phase locking steps more pronounced. For
$\kappa=1.5$, the dependence $\langle v \rangle (E_1)$ almost
coincides with the one for $\kappa=2$ (for larger amplitudes,
however, the three resonant steps $k=l=1$, $k=3,~l=2$, and $k=2,~l=1$
become much more visible).

Further decrease of the coupling constant makes the dependence
even less monotonic: for $\kappa=1$ (shown by $\diamond$ in the
figure), the non-monotonicity is quite weak, with a small interval
$0.145 < E_1 <0.154$ in which the kink velocity drops back to
zero. For $E_1 \gtrsim 0.2$, the dependence is also non-monotonic
because of the interaction of the kink with vibrational modes
localized on it. For $\kappa=0.5$ and $\kappa=0.75$, the average
kink velocity becomes strongly non-monotonic.

Notice from the right panel of Fig.~\ref{fig8} that several
reversals of the kink motion along the $E_1$ axis occur. In this
regard, we remark that the sign of $\langle v \rangle$ depends on
the relations between $\omega$, $\alpha$, and $\omega_{PN}$. Decrease 
of $\kappa$ can change $\omega_{PN}$ significantly [see
Eq. (\ref{PNfreq})], thus effecting the sign of the velocity.  The
reversals of $\langle v \rangle$ as a function of $E_1$, however,
are not fully explained by these arguments and we believe that, 
in analogy
with the continuous case, the coupling of the kink with small 
amplitude waves
plays also an important role.

By comparing the left and the right panels of Fig.~\ref{fig8}, one
can also see that the non-monotonic jumps occur in correspondence
with the transitions from chaotic to regular (phase locked)
regimes. In order to get a better understanding of this behavior,
we have plotted in insets (a) and (b) the mean velocity as a
function of the phase difference $\theta$. The inset (a) refers to
the case $\kappa=0.75$ for two different values of the driver
amplitude. We see that while for $E_1=0.19$ the dependence
$\langle v \rangle(\theta)$ is piecewise, but somewhat similar to
Fig.~\ref{fig2} (i.e., it has one minimum and one maximum), 
the dependence becomes more irregular as $E_1$ increases, with small
islands of phase locked regimes with opposite velocities. A similar
phenomenon  is observed, when $\kappa$ decreases to $\kappa=0.5$
[inset (b)]. This behavior is linked  to the well known phenomenon
of {\it crises of attractors}, i.e., to sudden appearance (or
disappearance) of an attractor as a system parameter varies
(see \cite{schuster88}).  By further decreasing the coupling
constant, more and more attractor crises are found.

From the insets of Fig.~\ref{fig8}, we also see that, in contrast
with the case $\kappa=1$ in Fig.~\ref{fig2},  the dependence
$\langle v \rangle=\langle v \rangle(\theta)$ loses any
resemblance with a sinusoidal function for small values of
$\kappa$. From Fig.~\ref{fig8}, it is also evident that the
depinning amplitude of the kink motion increases when $\kappa$ 
decreases. For very small $\kappa$, e.g., $\kappa=0.0625$ (not shown
in Fig.~\ref{fig8}), no directed motion is found - the kink is
either pinned by the lattice or destroyed by chaos when the
driving amplitude becomes large enough. This behavior is not
surprising, since we know that high discreteness normally prevents
solitons from free propagation.
%
%
\begin{figure}[htb]
\vspace{2pt}
\centerline{\epsfig{file=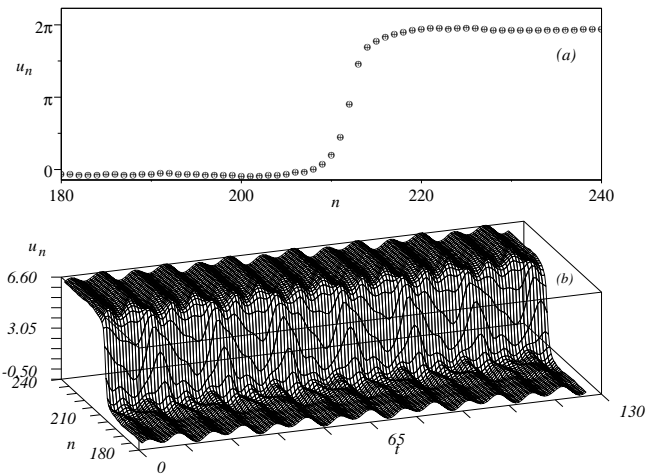,width=2.9in,angle=-90}}
\caption{Evolution of the kink profile $u_n(t)$. Upper panel
shows initial kink profile ($+$) and kink profile after
integration time $7T$ ($\circ$), shifted two sites backwards.
Parameters of the model are: $E_1=E_2=0.135$, $\theta=1.5$,
$\alpha=0.1$, $\omega=0.35$, and $\kappa=1$.} 
\label{l1}
\end{figure}

We remark that the attractor crises leading to reversal phenomena are
out of the range of validity of the  single particle approximation
since they involve many degrees of freedom. In particular, in the
limit of weak coupling, the point particle approximation (\ref{PN})
as well as the symmetry analysis discussed in Sec. II become not valid.

\subsection{Internal mode mechanism and dependence on the damping
parameter}
\label{imm}

In the case of continuous soliton ratchets, it was shown (see 
Refs.~\cite{SQ02,sz02pre,molinaPRL03,willis04,QSRS05}) that a
contribution to the ratchet phenomenon comes also from the
internal oscillation of the kink via the internal mode mechanism.
We expect this effect to be true also in the present case, 
both for large values (i.e., close to the continuum
limit) and intermediate and small values of the coupling constant 
$\kappa$ (for very
small values of $\kappa$, however, the kink becomes pinned to the
lattice and the ratchet phenomenon disappears as discussed
above).

In the following we investigate the internal mode mechanism by
fixing $\kappa=1$ and  performing direct numerical simulations of
Eqs.~(\ref{1}). In particular, we show the existence of a local
oscillation on the kink profile which is, perfectly synchronized
(phase locked) with the kink motion. We have found that when the
dynamics of the kink center of mass is phase locked to the resonance
$(k,l)$, the internal mode oscillation is also locked to the same
resonance. This is shown in Fig.~\ref{l1} for the discrete kink
ratchet, which is phase locked to the external driver with rotation
numbers $k=2,~l=7$. In the panel (a) of this figure, the initial and
final kink profiles are depicted, from which one can conclude that 
these profiles
perfectly coincide after $l=7$ periods of the external driver (notice
that the final configuration has been shifted by two sites backwards
in order to demonstrate full coincidence).
%
%
\begin{figure}[htb]
\vspace{2pt}
\centerline{\epsfig{file=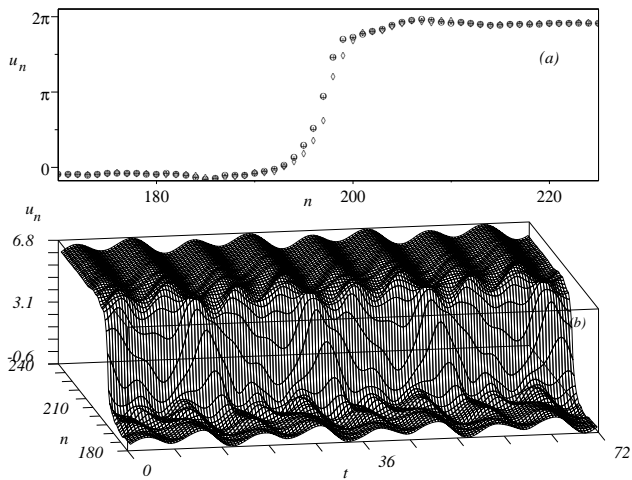,width=2.9in,angle=-90}}
\caption{Same as in Fig.~\ref{l1}. All parameters are the same
except $E_1=E_2=0.17$. In panel (a), the initial kink
configuration is shown by ($+$), the configuration after time
$T$ ($\diamond$) and after time $2T$ ($\circ$), shifted by one
site backwards. Panel (b) shows three-dimensional 
dynamical picture for the kink
evolution in the time interval $[0,4T]$.} 
\label{l2}
\end{figure}

The phase locking of the motion to the external driver is also
observed from the three-dimensional plot of $u_n(t)$ in panel (b).
Similar results are obtained for different values of the driving
amplitudes, as one can see from Fig.~\ref{l2}. In this case, the
kink is locked on the resonance with rotation numbers $(1,2)$
(notice that the kink reproduces itself completely after the
time $2T$). In panel (b), the complete dynamics in the time
interval $[0,4T]$ is also shown. The fact that the center of mass
motion and the oscillations on the kink profile (internal mode)
are perfectly synchronized suggests the existence of coupling
between the internal and translational modes similarly as for the
continuous case.

The existence of the internal mode mechanism in the discrete case
is also supported by the influence of the damping constant on the
phenomenon. In the continuous case, it was shown that the coupling
between the translational and internal modes decreases with the
damping and for the case of asymmetric potentials with symmetric
drivers, it was proved that in the limit $\alpha\rightarrow 0$, the
coupling completely disappears \cite{QSRS05}. In the case of
biharmonic asymmetric forces, the nonlinearity induces an effective
bias component, which gives rise to a point particle contribution
also in the absence of damping.
%
%
\begin{figure}[htb]
\vspace{2pt}
\centerline{\epsfig{file=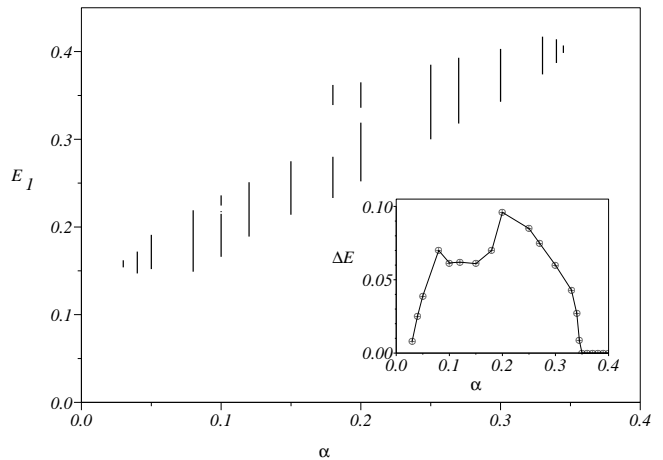,width=3in,angle=-90}}
\caption{Existence diagram of the main resonance ($k=l=1$) on the
plane ($\alpha, E_1=E_2$) for $\omega=0.35$, $\theta=2$, and
$\kappa=1$. The inset shows the dependence of the width of 
resonance $\Delta E$ on damping coefficient $\alpha$. When the
resonance consists of several islands, $\Delta E$ is computed as a
sum of the widths of individual islands. } 
\label{l3}
\end{figure}
The contribution of the internal mode to the ratchet dynamics,
however, should be sensitive to the damping, if the coupling is
controlled by the damping and the unidirectional motion should
become less effective for small values of the damping constant. On
the other hand, when the damping in the system becomes too large,
the dynamics are strongly reduced (or stopped), so that a
non-monotonous behavior of transport with $\alpha$ is expected and
an optimal value of the damping which maximizes the transport 
should exist.

In the following numerical study we have taken the width of
resonance steps produced by the ratchet dynamics (see Fig.~\ref{fig4}) 
as a measure of the efficiency of the transport in the
system (in the experimental setting considered in 
Ref.~\cite{uckzs04prl}, these resonance widths correspond to 
the voltage
steps in the IV characteristic of a Josephson junction).

Figure \ref{l3} illustrates the width of the main resonance $(1,1)$
as a function  of $\alpha$ (similar behavior is observed also for
other resonances). It follows from this figure that the resonant steps 
are reduced when
$\alpha$ is reduced and tend to disappear as $\alpha \rightarrow 0$ 
(for small $\alpha$ numerical calculations become very
difficult due to longer transient times to settle on the
attractor). The same behavior is observed as $\alpha$ increases
beyond the value $\alpha\approx 0.35$. From this figure we conclude
that the effect becomes  maximum around an intermediate value of
damping with $\alpha \approx 0.2$.
%
\begin{figure}[htb]
\vspace{2pt} 
\centerline{\epsfig{file=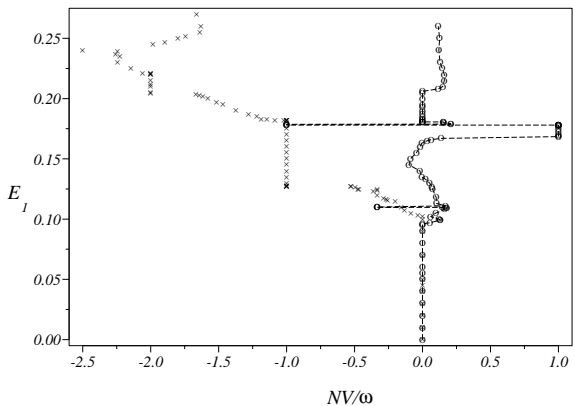,
width=2.7in,angle=-90}} 
\caption{Dependence $E_1=E_1(V)$ for
$\kappa=0.25$ ($\circ$) and $\kappa=1$ ($\times$). Other
parameters are: $\alpha=0.1$, $\theta=1.5$, and $\omega=0.35$. The
array consists of $N=10$ junctions. Dashed line for the case
$\kappa=0.25$ has been used as a guide for an eye.} 
\label{fig9}
\end{figure}
The existence of an optimal  coupling between the kink phase
looked dynamics and its internal oscillation which maximizes
the transport, indicates the validity of the internal mode mechanism
also for the discrete case.

\section{Applications to arrays of small Josephson junctions }
\label{jja}

In this section, we discuss possible applications of above
results to the case of an array of parallelly shunted and
AC-biased small Josephson junctions (JJAs). This system is
described by Eqs.~(\ref{1}) with $u_n$ corresponding to the phase
difference of the wave functions of the $n$th junction. The
discreteness parameter equals $\kappa=\sqrt{\Phi_0/(2\pi I_c L)}$,
where $\Phi_0$ is the magnetic flux quantum, $L$ is inductance of
an elementary cell, and $I_c$ is the critical current of an individual
junction. The dimensionless dissipation parameter is then
$\alpha=\Phi_0/(2\pi I_c R)$, where $R$ is the resistance of an
individual junction, and the time is normalized to the inverse 
Josephson
plasma frequency $1/\omega_0=\sqrt{C\Phi_0/(2\pi I_c)}$. In these
systems, the topological solitary waves have the physical meaning of
trapped magnetic flux quanta (fluxons) and  the voltage drop $V$
in the array is defined as 
\begin{equation}
 V= \frac{1}{N}\sum_{n=1}^N \lim_{t
\rightarrow \infty} \frac{1}{t}\int_0^t {\dot u}_n(t') dt' ~.
\end{equation}

The experiments with annular JJAs have been performed for
typical lengths $N \sim 8 \div 30$ (see 
Refs.~\cite{wzso96pd,u98pd,baufz00prl}). In the following, we consider
the case of an array with $N=10$ junctions subjected to periodic
boundary conditions (annular array). From the previous analysis we
expect that the ratchet dynamics of the fluxon give rise to
nonzero voltage drops in the IV characteristic of the array.
\begin{figure}[htb]
\vspace{2pt} \centerline{\epsfig{file=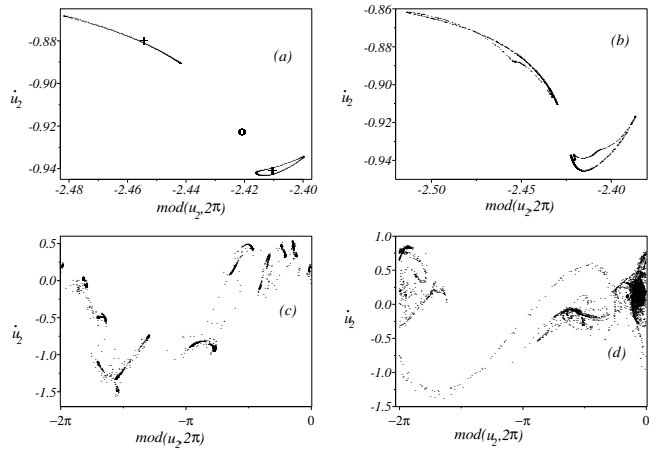,
width=2.7in,angle=-90}} 
\caption{Poincar\'e sections for
$\omega=0.35$, $\alpha=0.1$, $\theta=2$, and $N=10$ and different
values of driving amplitudes and coupling constants. Panels
(a)-(c) correspond to case $\kappa=1$. In panel (a),
$E_1=E_2=0.182$ ($\circ$), $E_1=E_2=0.18202$ ($+$),
$E_1=E_2=0.18205$ (dots). In panel (b), $E_1=E_2=0.18207$ and
in panel (c), $E_1=E_2=0.1821$. Panel (d) corresponds to
$\kappa=0.25$ and $E_1=E_2=0.097$.} 
\label{fig10}
\end{figure}

The voltage drop in an annular JJA (with one fluxon in it) is
related to the average fluxon velocity $\langle v \rangle$ by the
equation $V=2\pi \langle v \rangle /N$ \cite{u98pd}.

Figure \ref{fig9} shows the IV characteristic of an annular JJA
[dependence $E_1({V})$] for two different values of the coupling
constant $\kappa$. For $\kappa=1$ the characteristic is dominated
by large phase locking steps associated to limit cycles with
rotation numbers $(k,l)$. In particular, the phase locking resonances
($k=l=1$) and ($k=2,l=1$) are very well pronounced. On the limit
cycles of this type the voltage drop is equal to $V=k\omega/(Nl)$ 
and for each
lattice site $n$ we have $u_n(t+lT)=u_n(t)+2\pi k$ (notice that
for the phase locking dynamics, $NV/\omega$ corresponds to the
ratio of the rotation numbers $k/l$).

For smaller values of $\kappa$ (e.g., $\kappa=0.25$) directed fluxon
motion becomes diffusive almost for all values of the driving
amplitudes. This behavior corresponds to the irregular 
(non-vertical) parts
of the IV curve seen for $\kappa=0.25$. Notice that in this case,
only very few regular regimes are observed: pinning regions and
very small phase locking steps with $k=\pm 1,~l=1$. An idea of
complexity of the dynamics, resulting in this case, can be
obtained from the Poincar\'e sections of one of the junctions, as
reported in Fig.~\ref{fig10} for different values of the driver
amplitudes. The Poincar\'e section is produced by plotting the values
of the pair of the dynamical variables $\{u_n(t),{\dot u}_n(t)\}$
for $n=2$ after the time intervals $NT$. The phase locking regime
with the rotation numbers $k=l=1$ is identified as the fixed point
represented by the circle ($\circ$) in panel (a) of Fig.~\ref{fig10} 
and it corresponds to the fluxon which returns on the
initial site (junction) after one trip around the array during
the time $NT$. As the driving amplitude increases, the fixed
point bifurcates into a period-two-orbit corresponding to the two
crosses in the Poincar\'e section shown in Fig.~\ref{fig10}(a) 
(the kink dynamics
reproduces itself after the time $2NT$, making two trips around the
array, and so on). The curves shown in
panel (a) refer to the case $E_1=E_2=0.18205$ and correspond to
quasiperiodic motion. In panels (b) and (c) of Fig.~\ref{fig10}
we report the Poincar\'e sections of the period-doubling route to
chaos, while panel (d) shows the case with $\kappa=0.25$ for
which  chaos is fully developed and the fluxon motion is strongly
diffusive. A similar behavior is found for the Poincar\'e sections
taken on other junctions in the array.

Notice the presence  of voltage sign reversals in the
IV-characteristic in Fig.~\ref{fig9} generated by reversals of the
direction of the motion of the kink. In particular, the sudden
emergence of small intervals around $E_1 \simeq 0.178$ and $E_1
\simeq 0.11$, associated with negative voltages at the resonances
$k=-1,~l=1$ and $k=-1,~l=3$, respectively, are clearly visible.
These phenomena, as discussed in the previous section, are due to
attractor crises and have been observed also for other values
$\kappa$ as well as for longer arrays (see also Fig.~\ref{fig8}).
Comparing the dynamical behavior of arrays of different lengths
(different number of junctions), we find that already for a small
array of 10 junctions, the fluxon dynamics is qualitatively very similar
to that obtained for the infinite chain. The quantitative
difference, existing between the two cases, is probably due to the
interaction of the fluxon with the radiated small amplitude 
waves (Josephson plasmons),
this being  more pronounced in the discrete case due to
smaller size of the system.

\section{Conclusions}
\label{conc}

The ratchet phenomena induced by temporarily asymmetric zero mean fields 
on topological solitons (kinks and antikinks) of the
discrete sine-Gordon equation have been investigated. In
particular, we have studied the conditions for the occurrence of
discrete soliton ratchets and the dependence of the average
soliton velocity on the system parameters has been found. 
It has also being shown that, in
analogy with the continuous SG case, the unidirectional motion
arises, when all symmetries of the system relating orbits with
opposite velocities are broken. This condition can be achieved, 
using external biharmonic drivers of zero mean, which satisfy
suitable inequalities. The main characteristic of discrete soliton
ratchets, in comparison with the continuous case, is the existence
of a depinning threshold (in the driver amplitude) related to the
Peierls-Nabarro barrier. For driving amplitudes smaller than the
critical threshold, the kink performs periodic oscillations around
its center of mass, while above this threshold,the kink directed 
motion
takes place. The ratchet dynamics are effectively achieved, when
the kink motion is phase locked to the external driver. Besides
phase locking regimes, we have shown that the transport is possible
(but less effective) also in the presence of chaos and
intermittency. From this point of view, the phenomenon appears to
be much more complicated than the corresponding one in the
continuous SG system. The dependence of the mean velocity of the
kink on the amplitude, phase, and frequency of the biharmonic
driver, as well as on the damping  and on the coupling constant,
have extensively been investigated. We have also found that the
dependence of the velocity on the system parameters is much more
complicated in the discrete model, resembling in most cases a
devil's staircase. Some qualitative analogy with the continuous
case, however,  remains. In particular, the validity of the
internal mode mechanism for discrete ratchets has been
demonstrated.

Finally, the possibility to observe experimentally discrete
soliton ratchets in a one-dimensional  AC-biased annular array of small
Josephson junctions has been discussed in detail. From our results 
we predict that the kink ratchet dynamics induced by a biharmonic AC
driver of zero mean leads to the formation of voltage steps in
the IV characteristic of the array, in the absence of any DC bias.

\section{Acknowledgments}
\label{ack}

Y.Z. acknowledges the financial support from INTAS through its
Young Scientists Grant, contract number 03-55-1799. M.S.
acknowledges financial support from the MURST-PRIN-2005 grant
"Transport properties of classical and quantum systems".



\begin{thebibliography}{10}
\bibitem{note1} The effect should not be confused with the 
soliton dynamics under
the influence of a single harmonic sinusoidal signal in continuous
(see Ref.~\cite{gmu02pre}) or discrete 
(see Refs.~\cite{bm91prb,mfmfs97prb,fm99jpc}) media. 
In those cases, the soliton
motion becomes dependent on the initial phase and it 
can occur in both
directions with the same absolute value of the soliton velocity.
\end{thebibliography}
\end{document}